\documentclass[aps,twocolumn,superscriptaddress,amsmath,showpacs,tightenlines]{revtex4}
\usepackage{epsfig,graphicx,times}
\usepackage{amstext}
\usepackage{amsmath}            
\usepackage{amssymb}            
\usepackage{graphicx}           
\usepackage{latexsym}
\usepackage{bm}

\begin{document}
\title{Measuring the quality factor of a microwave cavity using superconduting qubit devices}

\author{ Yu-xi Liu}
\affiliation{Frontier Research System,  The Institute of Physical
and Chemical Research (RIKEN), Wako-shi 351-0198}
\author{L.F. Wei}
\affiliation{Frontier Research System,  The Institute of Physical
and Chemical Research (RIKEN), Wako-shi 351-0198}
\affiliation{Institute of Quantum Optics and Quantum Information,
Department of Physics, Shanghai Jiaotong University, Shanghai
200030, P.R. China }
\author{Franco Nori}
\affiliation{Frontier Research System,  The Institute of Physical
and Chemical Research (RIKEN), Wako-shi 351-0198}
\affiliation{Center for Theoretical Physics, Physics Department,
Center for the Study of Complex Systems, The University of
Michigan, Ann Arbor, Michigan 48109-1120}
\date{\today}

\begin{abstract}
We propose a method to create superpositions of two macroscopic
quantum states of a single-mode microwave cavity field interacting
with a superconducting charge qubit. The decoherence of such
superpositions can be determined by measuring either the Wigner
function of the cavity field or the charge qubit states. Then the
quality factor $Q$ of the cavity can be inferred from the
decoherence of the superposed states.  The proposed method is
experimentally realizable within current technology even when the
$Q$ value is relatively low, and the interaction between the qubit
and the cavity field is weak.

\end{abstract}

\maketitle \pagenumbering{arabic}

 \section{Introduction}

Superconducting (SC) Josephson junctions are considered promising
qubits for quantum information processing. This ``artificial
atom", with well-defined discrete energy levels, provides a
platform  to test fundamental quantum effects, e.g., cavity
quantum electrodynamics (QED). The study of the cavity QED of a SC
qubit, e.g., in Ref.~\cite{you}, can also open new directions for
studying the interaction between light and solid state quantum
devices. These can result in  novel controllable electro-optical
quantum devices in the microwave regime, such as microwave
single-photon generators and detectors. Cavity QED can allow the
transfer of information among SC qubits via photons, used as
information bus.

Recently, different information buses using bosonic systems, which
play a role analogous to a single-mode light field, have been
proposed to mediate the interaction between the SC qubits. These
bosonic ``information bus" systems can be modelled by:
nanomechanical resonators (e.g., in Refs.~\cite{nano}); large
junctions (e.g., Ref.~\cite{wang}); current-biased large junctions
(e.g., Refs.~\cite{large}), and  LC oscillators (e.g.,
Refs.~\cite{lc}). However, the enormous versatility provided by
photons should stimulate physicists to pay more attention to SC
qubits interacting via photons,while embedded inside a QED cavity.

Several theoretical proposals have analyzed the interaction
between SC qubits and quantized
~\cite{saidi,you,you1,liu,gao,vourdas,zagoskin} or classical
fields~\cite{zhou,paspalakis,liu1}. The strong coupling of a
single photon to a SC charge qubit has been experimentally
demonstrated~\cite{wallraff} by using a one-dimensional
transmission line resonator~\cite{blais}. But, the QED effect of
the SC qubit inside higher-dimensional cavities has not been
experimentally observed. The main roadblocks seem to be: i)
whether the cavity quality factor $Q$ can still be maintained high
enough when the SC qubit is placed inside the cavity. Different
from atoms, the effect of the SC qubit on the $Q$ value of the
cavity is not negligible due to its complex structure and larger
size. ii) The higher-dimensional QED cavity has relatively large
mode volume, making the interaction between the cavity field and
the qubit not be strong enough for the required quantum operations
within the decoherence time. iii) The transfer of information
among different SC qubits requires the qubit-photon interaction to
be switched on/off  by the external classical flux on time scales
of the inverse Josephson energy. A higher cavity $Q$ value, a
stronger qubit-photon interaction, and a faster switching
interaction for the SC qubit QED experiments, seem difficult to
achieve anytime soon.

In view of the above problems, it would be desirable to explore
the possibility to demonstrate a variety of relatively simple
cavity QED phenomena with a SC qubit. The determination of the
cavity $Q$ value is a very important first step for the
experiments on cavity QED with SC qubits. However, theoretical
calculations of the $Q$ value are not always easy to perform
because of the complexity of the circuit. Recent
experiments~\cite{pkd} on broadband SC detectors showed that the
$Q$ value of the SC device can reach $2\times 10^{6}$, which
indicates that relatively simple experiments using cavity QED with
a SC qubit are possible.

In this paper, we propose an experimentally feasible method which
can be used to demonstrate a simple cavity QED effect of the SC
qubit. For instance, superpositions of two macroscopic quantum
states of a single-mode microwave cavity field can be created by
the interaction between a SC charge qubit and the cavity field. At
this stage, the injected light field is initially a coherent
state, which can be easily prepared. The decoherence of the
created superposition states can be further determined by
measuring either the Wigner function of the cavity field or the
charge qubit states. Then the cavity $Q$ value can be inferred
from this decoherence measurement. Our proposal only needs few
operations with a relatively low Q value. Also, we do not need to
assume a very fast sweep rate of the external magnetic field for
switching on/off the qubit-field interaction. Furthermore, the
qubit-field interaction is not necessarily resonant.

We begin in Sec. II with a brief overview of the qubit-field
interaction. In Sec. III, we discuss how to prepare superpositions
of two different cavity field states under the condition of large
detuning. In Sec. IV, the cavity $Q$ value is determined by the
tomographic reconstruction of the cavity field Wigner function. In
Sec. V, we show an alternative method to determine the $Q$ value
according to the states of the qubit. Finally, we list our
conclusions.

\section{Theoretical model}
We briefly review a model of a SC charge qubit  inside a cavity.
The Hamiltonian can be written as~\cite{you,you1,liu,gao}
\begin{eqnarray}\label{eq:1}
&&H=\hbar\omega a^{\dagger}a+E_{z}\sigma_{z}\\
&&-\,E_{J}\sigma_{x} \cos\left[\frac{\pi}{\Phi_{0}}\left(\Phi_{c}
I+\eta \,a+\eta^{*}\,a^{\dagger}\right)\right],\nonumber
\end{eqnarray}
where the first two terms respectively represent the free
Hamiltonians of the cavity field with frequency $\omega$ for the
photon creation (annihilation) operator $a^{\dagger} \,(a)$, and
the qubit charging energy
\begin{equation}\label{charge}
E_{z}=-2E_{\rm ch}(1-2n_{g})\, ,
\end{equation}
which depends on the gate charge $n_{g}$. The single-electron
charging energy is $E_{\rm ch}=e^2/2(C_{g}+2C_{J})$ with  the
capacitors $C_{g}$ and $C_{J}$ of the gate and the Josephson
junction, respectively. The dimensionless gate charge,
$n_{g}=C_{g}V_{g}/2e$, is controlled by the gate voltage $V_{g}$.
Here, $\sigma_{z}$, $\sigma_{x}$ are the Pauli operators, and the
charge excited state $|e\rangle$ and ground state $|g\rangle$
correspond to the eigenstates $
|\!\downarrow\rangle=\left(\begin{array}{l}0\\1\end{array}\right)
$ and
$|\!\uparrow\rangle=\left(\begin{array}{l}1\\0\end{array}\right)
$
of the spin operator $\sigma_{z}$, respectively. $I$ is an
identity operator. The third term  is the nonlinear qubit-photon
interaction. $E_{J}$ is  the Josephson energy for a single
junction.  The parameter $\eta$ is defined as $\eta=\int_{S}
\mathbf{u}(\mathbf{r})\cdot d\mathbf{s}$ with the mode function of
the cavity field $\mathbf{u}(\mathbf{r})$, $S$ is the surface
defined by the contour of the SQUID.  We can decompose the cosine
in Eq.~(\ref{eq:1}) into classical and quantized parts. The
quantized parts $\sin[\pi(\eta\, a+H.c.)/\Phi_{0}]$ and
$\cos[\pi(\eta\, a+H.c.)/\Phi_{0}]$ can be further expanded as a
power series in $a \,(a^{\dagger})$. To estimate the qubit-photon
coupling constant, the qubit is assumed to be inside a full-wave
cavity  with the standing-wave form for a single-mode magnetic
field~\cite{scully}
\begin{equation}\label{eq:m1}
B_{x}=-i\sqrt{\frac{\hbar\omega}{\varepsilon_{0}V
c^{2}}}(a-a^{\dagger})\cos(k z).
\end{equation}
The polarization of the magnetic field is along the normal
direction of the surface area of the SQUID, located at an antinode
of the standing-wave mode. The mode function
$\sqrt{\hbar\omega/\varepsilon_{0}V c^{2}}\cos(k z)$ can be
assumed to be independent of the integral area because the maximum
linear dimension of the SQUID, e.g., even for $50 \,\,\mu$m,  is
much less than $0.1$ cm, the shortest microwave wavelength of the
cavity field. Then, in the microwave regime, the estimated range
of values for $\pi\eta/\Phi_{0}$ is: $8.55 \times 10^{-6}\leq
\pi\eta/\Phi_{0}\leq 1.9\times 10^{-3}$, for a fixed area of the
SQUID, e.g., $50 \,\mu$m $\times 50\, \mu$m. If the light field is
not so strong  (e.g., the average number of photons inside the
cavity $N=\langle a^{\dagger}a\rangle\leq 100$), then we can only
keep the first order of $\pi\eta/\Phi_{0}$ and safely neglect all
higher orders. Thus, the Hamiltonian~(\ref{eq:1}) becomes
\begin{eqnarray}\label{eq:2}
&&H=\hbar\omega a^{\dagger}a+E_{z}\sigma_{z}-\,E_{J}\sigma_{x}\cos
(\frac{\pi\Phi_{c}}{\Phi_{0}})\nonumber \\
&&+\frac{\pi E_{J}}{\Phi_{0}}\sin (\frac{\pi\Phi_{c}}{\Phi_{0}})
\left(\eta
\,a\,\sigma_{+}+\eta^{*}\,a^{\dagger}\,\sigma_{-}\right).
\end{eqnarray}
It is clear that the qubit-photon interaction can be controlled by
the classical flux $\Phi_{c}$, after neglecting  higher-orders in
$\pi\eta/\Phi_{0}$.

\section{Preparation of macroscopic superposition states}

The qubit-photon system can be initialized  by adjusting the gate
voltage $V_{g}$ and the external flux $\Phi_{c}$ such that
$n_{g}=1/2$ and $\Phi_{c}=0$, then the dynamics of the qubit-field
is governed by the Hamiltonian
\begin{equation}\label{eq:3}
H_{1}=\hbar\omega a^{\dagger}a-E_{J}\sigma_{x}.
\end{equation}
Now there is no interaction between the cavity field and the
qubit; thus, the cavity field and the qubit evolve according to
Eq.~(\ref{eq:3}). We assume that the qubit-photon system works at
low temperatures $T$ ( e.g., $T=30$ mK in Ref.~\cite{nakamura}),
then the mean number of thermal photons $\langle n_{th}\rangle$ in
the cavity can be negligible in the microwave regime~\cite{liu},
and the cavity is approximately considered in the zero temperature
environment. The initial state of the cavity field is prepared by
injecting a single-mode coherent light
\begin{equation}
|\alpha\rangle=\exp\left\{-\frac{|\alpha|^2}{2}\right\}\sum_{n=0}\frac{\alpha^{n}}{\sqrt{n!}}|n\rangle\,,
\end{equation}
into the cavity. Here, without loss of generality, $\alpha$ is
assumed to be a real number, and
$a|\alpha\rangle=\alpha|\alpha\rangle$. The qubit is assumed to be
initially in the ground state $|g\rangle$. After a time interval
$\tau_{1}=\hbar\pi/4E_{J}$, the qubit ground state $|g\rangle$ is
transformed as $|g\rangle\rightarrow
\left(|g\rangle+i|e\rangle\right)/\sqrt{2}$; then, the
qubit-photon state evolves into
\begin{equation}\label{eq:4}
|\psi(\tau_{1})\rangle=
\frac{1}{\sqrt{2}}(|g\rangle+i|e\rangle)|\alpha\rangle\, .
\end{equation}
Here we have neglected the free evolution phase factor
$e^{-i\omega \tau_{1}}$ in $\alpha$.

Now, we assume that the gate voltage and the magnetic flux are
switched to $n_{g}\neq 1/2$ (this value of $n_{g}$ will be
specified later) and $\Phi_{c}=\Phi_{0}/2$, respectively. Then the
qubit-photon interaction appears and the effective Hamiltonian
governing the dynamic evolution of the qubit-photon can be written
as (see Appendix A)
\begin{equation}\label{eq:5}
H_{2}=\hbar\omega_{-} a^{\dagger }a+\frac{1}{2}\hbar \Omega
\sigma_{z}+\hbar\frac{|g|^{2}}{\Delta }\left(1+2 a^{\dag }a\right)
|e\rangle \langle e|\, ,
\end{equation}
with $\omega_{-}=\omega -|g|^{2}/\Delta $ and $ g=(\pi\eta
E_{J})/(\hbar{\Phi_{0}})$. The detuning $\Delta=\Omega-\omega > 0$
between the qubit transition frequency
$\Omega=-4E_{ch}(1-2n_{g})/\hbar$ and the cavity field frequency
$\omega$ is assumed to satisfy the large detuning condition
\begin{equation}\label{large}
\frac{\pi E_{J}|\eta|}{\hbar\Phi_{0}\Delta}\ll 1.
\end{equation}
The unitary evolution operator corresponding to Eq.~(\ref{eq:5})
can be written as
\begin{eqnarray}\label{eq:6}
U(t)&=&\exp\left[-i\left(\omega_{-} a^{\dagger}a+\frac{\Omega}{2}
\sigma_{z}\right)t\right] \nonumber\\
&\times&\exp\left[-it F (a^{\dagger}a)|e\rangle \langle e|\right],
\end{eqnarray}
here, the operator $F(a^{\dagger}a)$ is expressed as
\begin{equation}
F(a^{\dagger}a)=\frac{|g|^2}{\Delta}(1+2 a^{\dagger}a).
\end{equation}
With an evolution time $\tau_{2}$, the state (\ref{eq:4}) evolves
into
\begin{equation}\label{eq:8}
|\psi(\tau_{2})\rangle=\frac{1}{\sqrt{2}}\left[|g\rangle|\beta\rangle+
i\exp(i\theta)|e\rangle|\beta^{\prime}\rangle\right]
\end{equation}
where a global phase $\exp(-i\Omega \tau_{2}/2)$ has been
neglected, $\theta=(\Omega-|g|^2/\Delta)\tau_{2}$, $\beta=\alpha
\exp[-i\omega_{-} \tau_{2}]$, and $\beta^{\prime}=\beta
\exp(-i\phi)$, with $\phi=2|g|^2 \tau_{2}/\Delta$.
Equation~(\ref{eq:8}) shows that a phase shift $\phi$ is generated
for the coherent state $|\beta\rangle$ of the cavity field when
the qubit is in the excited state $|e\rangle$, but the qubit
ground state $|g\rangle$ does not induce an extra phase for the
coherent state $|\beta\rangle$.

The gate voltage and the magnetic field are now adjusted such that
the conditions $n_{g}=1/2$ and $\Phi_{c}=0$ are satisfied; then
the qubit-photon interaction is switched off. Now let the system
evolve a time $\tau^{\prime}=\tau_{1}=\hbar\pi/4E_{J}$, then
Eq.~(\ref{eq:8}) becomes
\begin{eqnarray}\label{eq:9}
|\psi(\tau_{2})\rangle&=&\frac{1}{2}|g\rangle\otimes[|\beta\rangle-\exp(i\theta)|\beta^{\prime}\rangle]\nonumber\\
&+&i\frac{1}{2}|e\rangle\otimes[|\beta\rangle+\exp(i\theta)|\beta^{\prime}\rangle],
\end{eqnarray}
where a free phase factor $e^{-i\omega \tau_{1}}$ in the cavity
field states $|\beta\rangle$ and  $|\beta e^{-i\phi}\rangle$ has
been neglected.

The superpositions of two distinguished coherent states can be
conditionally generated by measuring the charge states of the
qubit as,
\begin{equation}\label{eq:10}
|\beta_{\pm}\rangle=N^{-1}_{\pm}[|\beta\rangle\pm
\exp(i\theta)|\beta^{\prime}\rangle],
\end{equation}
where the $+$ ($-$) correspond to the measurement results
$|e\rangle$ ($|g\rangle$), and the normalized constants $N_{\pm}$
are determined by
\begin{equation}
N^{2}_{\pm}=2\pm
2\cos\theta^{\prime}\exp\!\left[-2|\alpha|^2\sin^2\left(\frac{\phi}{2}\right)\right],
\end{equation}
where $\theta^{\prime}=|\alpha|^2\sin\phi-\theta$, and the
relation $|\beta|^2=|\alpha|^2$ is used.

Due to $\Phi_{c}=0$, after the superpositions in Eq.(\ref{eq:10})
are created, the dynamic evolution of the cavity field is only
affected by its dissipation, characterized by the decay rate
$\gamma$, which can be expressed by virtue of the cavity quality
factor $Q$ as $Q=\omega/\gamma$. Now let the cavity field
described by the states $|\beta_{+}\rangle$ or $|\beta_{-}\rangle$
evolve a time $\tau_{3}$; then the reduced density matrices of the
superpositions can be described by
\begin{eqnarray}\label{eq:11}
&&\rho_{\pm}(\tau_{3}) =\frac{1}{N^{2}_{\pm}}\left\{|\beta
u\rangle\langle \beta u|+|\beta^{\prime} u \rangle\langle
\beta^{\prime} u |\right.\nonumber \\
&&\left.\pm  C  |\beta^{\prime} u \rangle\langle \beta u |\pm
C^{*}|\beta u \rangle\langle \beta^{\prime} u |\right\},
\end{eqnarray}
where
\begin{equation}
C=\exp(i\theta) \exp\left\{|\alpha|^2(1-e^{-i\phi})(u^2-1)\right\}
\end{equation}
 and
\begin{equation}\label{eq:188}
u\equiv u(\tau_{3})=\exp\left(-\frac{\gamma}{2} \tau_{3}\right)
=\exp\left(-\,\frac{\omega\tau_{3}}{2Q} \right).
\end{equation}
It is clearly shown that the mixed state in Eq.~(\ref{eq:11}) is
strongly affected by the $Q$ value. Equation~(\ref{eq:11}) is
derived for zero temperature since thermal photons are negligible
at low-temperature. Equations~(\ref{eq:11}-\ref{eq:188}) show that
the information of the cavity quality factor $Q$ can be encoded in
a reduced density matrix of the cavity field. The $Q$ value can be
determined using two different methods, after encoding its
information in Eqs.(\ref{eq:11}-\ref{eq:188}). Below, we will
discuss these two approaches.

\section{Measuring $Q$ by photon state tomography}
\begin{figure}
\includegraphics[width=42mm]{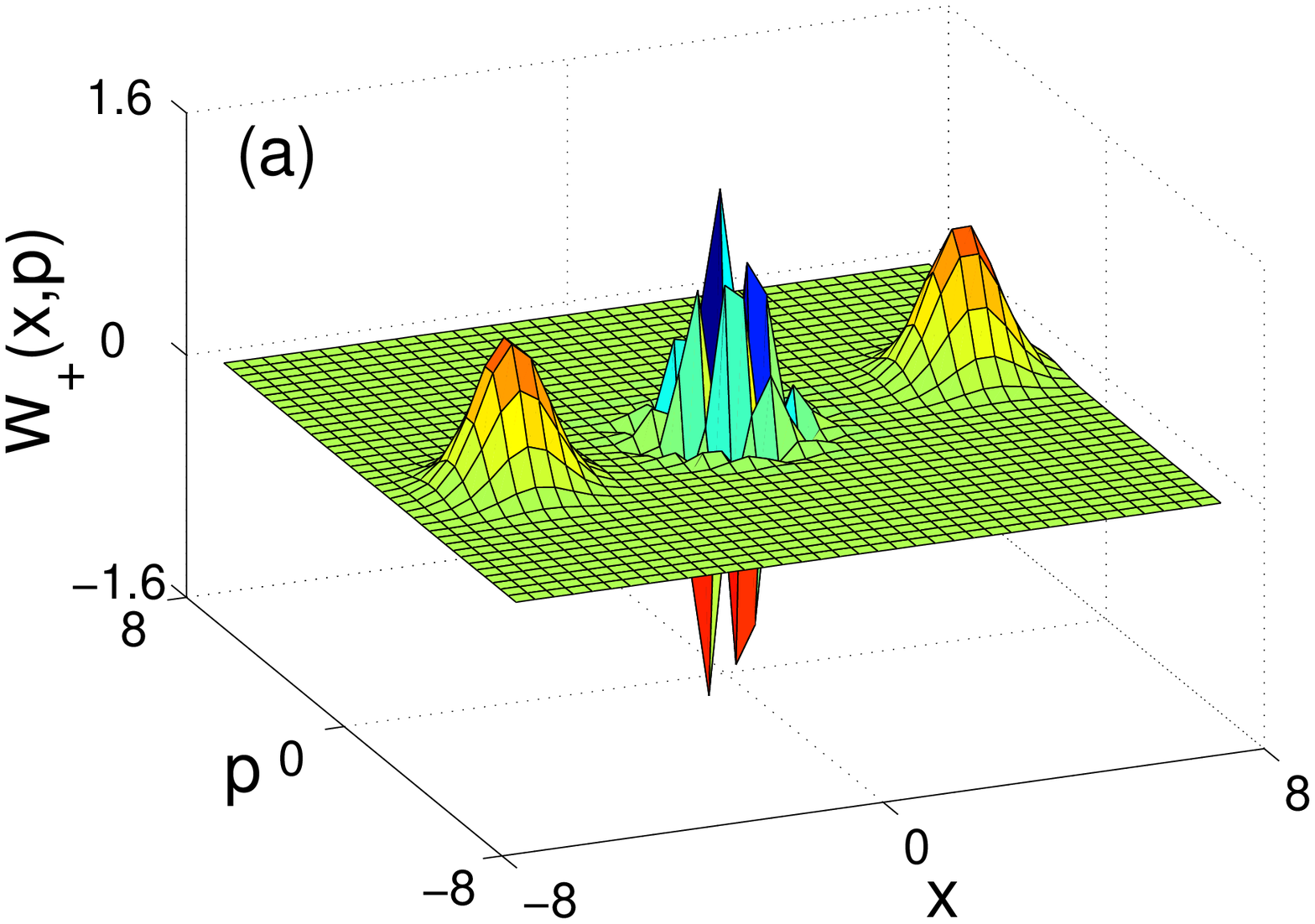}
\includegraphics[width=42mm]{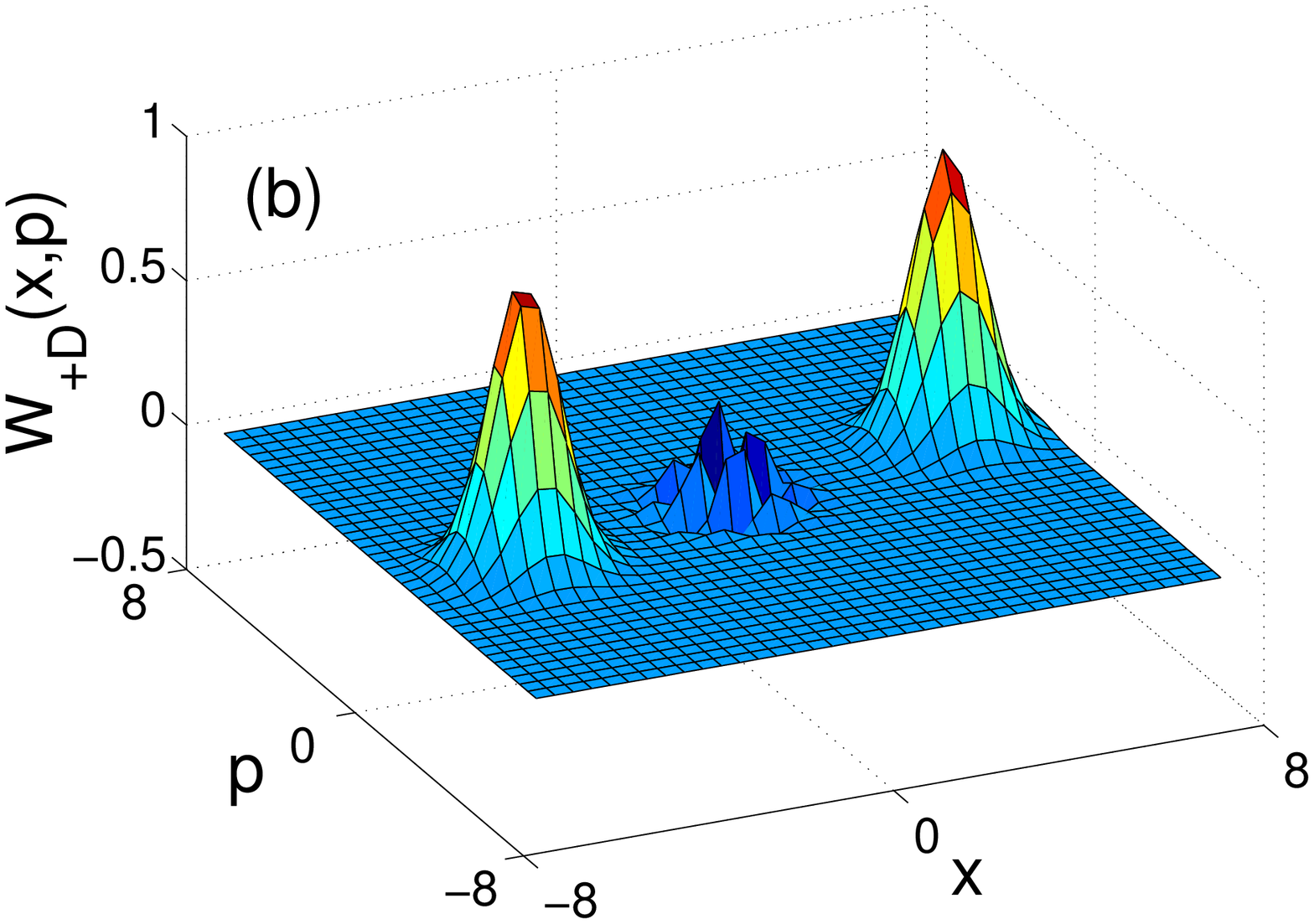}
\caption[]{(Color online) Wigner functions $W_{+}(x,p)$ of
Eq.~(\ref{eq:10}) and $W_{+D}(x,p)$ of Eq.~(\ref{eq:11}) for the
cavity field without and with the energy dissipation are shown in
(a) and (b), for the input state $|\alpha\rangle$ with
$\overline{n}=16$. Here the Wigner functions $W_{+}(x,p)$ and
$W_{+D}(x,p)$ are normalized to $\pi N_{+}$.}\label{fig1}
\end{figure}

The state $\rho$ of the optical field is generally visualized when
it is represented by a Wigner function~\cite{hans} in the position
$x$ and momentum $p$ space, which is  written as
\begin{equation}
W(x,p)=\frac{1}{\pi}\int_{-\infty}^{\infty}\langle
x-x^{\prime}|\rho|x+x^{\prime}\rangle e^{i2p  x^{\prime}}{\rm d}
x^{\prime}.
\end{equation}
The Wigner function $W(x,p)$ can be experimentally measured by
state tomographic techniques~\cite{hans}. For any two coherent
states, $|\alpha\rangle$ and $|\beta\rangle$, the Wigner function
$W(x,p)$ can be represented as
\begin{eqnarray}\label{eq:17}
W_{\alpha,\beta}(x,p)&=&\frac{1}{\pi}\int_{-\infty}^{\infty}\langle
x-x^{\prime}|\alpha\rangle\langle\beta|x+x^{\prime}\rangle e^{i2p
x^{\prime}}{\rm d} x^{\prime} \nonumber\\
&=&\frac{1}{\pi}\exp\left\{-\frac{1}{2}(|\alpha|^2+|\beta|^2-2\alpha\beta^*)\right\}\nonumber\\
&\times&\exp\left\{-(x-q_{1})^2-(p+iq_{2})^2\right\}
\end{eqnarray}
with $q_{1}=(\alpha+\beta^*)/\sqrt{2}$ and
$q_{2}=(\alpha-\beta^*)/\sqrt{2}$. The Wigner functions
$W_{\pm}(x,p)$ and $W_{\pm D}(x,p)$ for the states (\ref{eq:10})
and (\ref{eq:11}) were calculated (see Appendix B) by using
Eq.~(\ref{eq:17}). Comparing the tomographically measured results
for the states (\ref{eq:10}) and (\ref{eq:11}), the $Q$ factor of
the cavity can be finally determined, as explained below by using
an example.

We further numerically calculate the Wigner functions
$W_{\pm}(x,p)$ and $W_{\pm D}(x,p)$ of the states (\ref{eq:10})
and (\ref{eq:11}) from the SC qubit parameters and given operation
durations. Using  current values for experimental data, the basic
physical parameters can be specified as follows. We assume that
the SC Cooper-pair box is made from aluminum, with a BCS energy
gap of $\sim 2.4$K (about 50 GHz)~\cite{lehnert}, the charge
energy $E_{\rm ch}$ and the Josephson energy $E_{\rm J}$ are
$4E_{\rm ch}/h=149$ GHz and $2E_{\rm J}/h=13.0$ GHz,
respectively~\cite{lehnert}. The frequency of the cavity field is
taken as $40$ GHz, corresponding to a wavelength $\sim 0.75$ cm.
The above numbers show that the SC energy gap is the largest
energy, so the quasi-particle excitation on the island can be well
suppressed at low temperatures, e.g., $20$ mK . The SQUID area is
assumed to be about $50 \,\mu$m $\times 50\, \mu$m, then the
absolute value $|g|$ of the qubit-photon coupling constant is
about $|g|=4\times 10^{6}$ rad s$^{-1}$.

Let us now prepare entangled qubit-photon states as in
Eq.~(\ref{eq:9}). Any gate charge value $n_{g}$ in
Eq.~(\ref{charge}), in which the large detuning condition in
Eq.~(\ref{large}) is satisfied, can be chosen to realize our
proposal. For concreteness, we give an example. The gate voltage
is adjusted such that the gate charge is $n_{g}\approx 0.634233$,
which can be experimentally achieved~\cite{lehnert}, then the
detuning $\Delta=\Omega-\omega\approx 9.0\times 10^{6}$ rad
s$^{-1}$. Thus, $\Omega$ is about $40$ GHz plus $1.4$ MHz, and
$|g|^2/\Delta \,\simeq\, 0.27$ MHz. We can also find that
$\Delta/|g|\approx 2.3$, so a large-detuning condition can be
used~\cite{xm,pt}.  For a given Josephson energy,  $2E_{J}/h=13.0$
GHz, the operation times $\tau_{1}=4.8\times 10^{-12}$ s, required
to prepare a superposition of $|e\rangle$ and $|g\rangle$ with
equal probabilities, is much less than the qubit relaxation time
$T_{1}=1.3\,\mu$s and dephasing time $T_{2}=5$ ns.  We can choose
the duration $\tau_{2}$ for a given input coherent state
$|\alpha\rangle$ with the condition,  that the distance
$|\beta-\beta^{\prime}|$ between two coherent states
$|\beta\rangle$ and $|\beta^{\prime}\rangle$ satisfies
\begin{equation}
|\beta-\beta^{\prime}|=2|\alpha|
\sin\left(\frac{\phi}{2}\right)>1.
\end{equation}
So the lower bound of the duration $\tau_{2}$ can be given as
\begin{equation}\label{18}
\tau_{2}=\frac{\Delta}{|g|^2}\arcsin\left(\frac{1}{2|\alpha|}\right)\,,
\end{equation}
when $0 \leq \phi\leq \pi$. Equation~(\ref{18}) shows that a
shorter $\tau_{2}$ can be obtained for a higher intensity
$|\alpha|$ with fixed detuning $\Delta$ and coupling constant $g$.

As an example, we plot the Wigner function of the superposition
$|\beta_{+}\rangle$ in Fig.~\ref{fig1}(a) for an input coherent
light $|\alpha\rangle$ with the mean photon number
$\overline{n}=|\alpha|^2=16$.  We choose a simple case $\phi=\pi$,
corresponding to the operation time $\tau_{2}\approx 0.93 \,\mu$s,
which is less than the qubit lifetime $T_{1}$ and the cavity field
lifetime $T_{\rm ph}\approx 2\,\mu$s for a bad cavity with
$Q=5\times 10^{5}$. In such a case, $\beta^{\prime}=-\beta$ and
the phase $\theta$ is about $0.996\,[{\rm mod}\, 2\pi]$ rad. Other
parameters used in Fig.~\ref{fig1} are given above. If we set the
evolution time $\tau_{3}=0.1\, \mu$s, then the Wigner function of
Eq.~(\ref{eq:11}) for the above cavity quality factor is shown in
Fig.~\ref{fig1}(b). The central structure in Fig.~\ref{fig1}(a)
represents the coherence of the quantum state. In
Fig.~\ref{fig1}(b), we find that the height of the Wigner function
$W_{+D}(x,p)$, especially for the central structure, is reduced by
the environment. Comparing Fig.~\ref{fig1}(a) and
Fig.~\ref{fig1}(b), it is found that the coherence of the
superposed states is suppressed by the environment, and the
decoherence of superpositions is tied to the energy dissipation of
the cavity field. Then, the $Q$ value can in principle be
estimated  by measuring the Wigner functions of Eqs.~(\ref{eq:11})
and (\ref{eq:10}), and comparing these two kinds of results.

\section{Determining $Q$ by readout of charge states}

The determination of the $Q$ value by measuring the Wigner
function needs optical instruments. In solid state experiments,
the charge states are typically measured.  Instead of using
optical instruments, it would be desirable to obtain the $Q$ value
by measuring charge states. This will be our goal here. The
process to achieve this can be described as follows.

i) According to the measurements on the charge qubit states in
Eq.(\ref{eq:9}), the qubit-photon states are projected to
$|g\rangle\otimes|\beta_{-}\rangle$ or
$|e\rangle\otimes|\beta_{+}\rangle$, respectively. After the
evolution time $\tau^{\prime}_{3}$, a $\pi/2$ quantum operation is
performed on the qubit with the duration $T=\hbar\pi/4E_{J}$.
Then, the qubit ground state $|g\rangle$, or excited state
$|e\rangle$, is transformed into the superposition
$(|g\rangle+i|e\rangle)/\sqrt{2}$, or
$(i|g\rangle+|e\rangle)/\sqrt{2}$, and the photon states
$|\beta_{\pm}\rangle$ evolve into mixed states after the evolution
time $\tau=\tau^{\prime}_{3}+T$, and the photon-qubit states can
be expressed as
\begin{equation}\label{eq:18}
\rho_{Q+F}=\frac{1}{2}(|g\rangle\pm i|e\rangle)(\langle g|\mp
i\langle e|)\otimes \rho_{\pm}(\tau)\, ,
\end{equation}
with subscripts $Q$ and $F$ denoting the qubit and cavity field,
respectively.  The reduced density matrices $\rho_{\pm}(\tau)$
take the same form as in Eq.~(\ref{eq:11}) with $\tau$ replacing
$\tau_{3}$.

ii) After the above procedure, the qubit-photon interaction is
switched on by applying the external magnetic flux
$\Phi_{e}=\Phi_{0}/2$. By using Eq.~(\ref{eq:6}),
Eq.~(\ref{eq:18}) evolves into
\begin{eqnarray}\label{eq:19}
2\rho_{A+F}^{(1)} &=&|g\rangle\langle g|\otimes
U_{1}(\tau_{4})\rho_{\pm}(\tau)U^{\dagger}_{1}(\tau_{4})\nonumber\\
&+&|e\rangle\langle e|\otimes
U_{2}(\tau_{4})\rho_{\pm}(\tau)U^{\dagger}_{2}(\tau_{4}) \\
&\mp& i\exp(-i\Omega_{-}\tau_{4})|g\rangle\langle e|\otimes
U_{1}(\tau_{4})\rho_{\pm}(\tau)U^{\dagger}_{2}(\tau_{4})\nonumber\\
&\pm&i\exp(+i\Omega_{-}\tau_{4})|e\rangle\langle g|\otimes
U_{2}(\tau_{4})\rho_{\pm}(\tau)U^{\dagger}_{1}(\tau_{4})\nonumber
\end{eqnarray}
with $\Omega_{-}=\Omega-|g|^2/\Delta$, and a shorter evolution
time $\tau_{4}$.  For example, $\tau_{4}$ is less than the
lifetime $T_{1}$ of the qubit at least.  The time evolution
operators $U_{1}(\tau_{4})$ and $U_{2}(\tau_{4})$ in
Eq.~(\ref{eq:19}) are
\begin{subequations}
\begin{eqnarray}
U_{1}(\tau_{4})&=&\exp\left[-i \omega_{-}\,a^{\dagger}a\,\tau_{4}\right],\\
U_{2}(\tau_{4})&=&\exp\left[-i\omega_{+}\,a^{\dagger}a\,\tau_{4}\right].
\end{eqnarray}
\end{subequations}
with $\omega_{\pm}=\omega\pm|g|^2/\Delta$. After this qubit-photon
interaction, the information of the $Q$ value is encoded.

 iii) The qubit-photon coupling is switched off and a $\pi/2$
rotation is made on the qubit. If the state of the cavity field is
prepared to $|\beta_{-}\rangle$ of Eq.~(\ref{eq:10}) in the first
step, then the qubit is in the ground state $|g\rangle$. After
measuring the qubit states, the photon states are projected to
\begin{subequations}
\begin{equation}\label{eq:21}
\rho_{e/g}=\frac{1}{4}(A \pm B)
\end{equation}
where the sign $``+"$ corresponds to the excited state $|e\rangle$
measurement, but $``-"$ corresponds to the ground state
$|g\rangle$ measurement. The operators $A$ and $B$ are
\begin{eqnarray}
A&=&\sum_{i=1}^{2}U_{i}(\tau_{4})\rho_{-}(\tau)U^{\dagger}_{i}(\tau_{4})\label{eq:21b},\\
B&=&2 \,{\rm
Re}[\exp(-i\Omega_{-}\tau_{4})U_{1}(\tau_{4})\rho_{-}(\tau)U^{\dagger}_{2}(\tau_{4})]\label{eq:21c}.
\end{eqnarray}
\end{subequations}

After tracing out the cavity field state, the probabilities
corresponding to measuring charge states $|e\rangle$ and
$|g\rangle$ are
\begin{eqnarray}\label{eq:22}
P_{e/g}(\tau)&=&{\rm Tr}_{F}\,\left(\rho_{e/g}\right)\nonumber\\
&=&\frac{1}{2}\left\{1\pm {\rm Re}({\rm
Tr}_{F}[\exp(-i\varphi)\rho_{-}(\tau)])\right\}
\end{eqnarray}
with $\varphi=(\Omega_{-}-2|g|^2 a^{\dagger}a/\Delta)\tau_{4}$.
Then the measurement probabilities are related to the $Q$ values.
Substituting $\rho_{-}(\tau)$ into Eq.(\ref{eq:22}), we can obtain
\begin{subequations}\label{eq:24}
\begin{eqnarray}
&&{\rm Re}\left\{{\rm Tr}[\exp(-i\varphi)\rho_{-}(\tau)]\right\} \\
&&=\frac{2}{N^2_{-}}\exp\left[-2\alpha(\tau)\sin^2\phi^{\prime}\right]
\cos\left[\Omega_{-}\tau_{4}-\alpha(\tau)\sin(2\phi^{\prime})\right]\nonumber\\
&&- \frac{1}{N^2_{-}}\cos\left[\theta_{-}
-|\alpha|^2\sin\phi+\theta-\Omega_{-}\tau_{4}\right]\exp\left(+G_{-}-\Gamma\right)\nonumber\\
&&- \frac{1}{N^2_{-}} \cos\left[\theta_{+}
+|\alpha|^2\sin\phi-\theta-\Omega_{-}\tau_{4}\right]\exp\left(-G_{+}-\Gamma\right)
\nonumber
\end{eqnarray}
with the parameters
\begin{eqnarray}
\phi^{\prime}&=&\frac{|g|^2}{\Delta}\tau_{4},\\
\Gamma&=&2|\alpha|^2\sin^2(\frac{\phi}{2}),\\
\alpha(\tau)&=&|\alpha u(\tau)|^2,\\
G_{\pm}&=&2\alpha(\tau)\sin\phi^{\prime}\sin(\phi\pm\phi^{\prime}),\\
\theta_{\pm}&=&2\alpha(\tau)\cos(\phi\pm
\phi^{\prime})\sin\phi^{\prime}.
\end{eqnarray}
\end{subequations}
From Eq.~(\ref{eq:24}), we find that $\phi^{\prime}$ should
satisfy the condition $\phi^{\prime}\neq n\pi$  for $\phi=\pi$, in
order to describe the dissipation effect; here $n$ is an integer.
Generally speaking, if one of the functions $G_{+}$, $G_{-}$,
$\theta_{+}$, $\theta_{-}$, $\sin\phi^{\prime}$, or
$\sin(2\phi^{\prime})$ is nonzero, then this is enough to encode
the $Q$ value, which can be obtained from Eq.~(\ref{eq:24}),
together with Eq.~(\ref{eq:188}), using $\tau$ instead of
$\tau_{3}$.

However, if the superposition of the cavity fields is prepared to
the state $|\beta_{+}\rangle$ in the first step, then the ground
and excited state measurements make the cavity field collapse to
state
\begin{equation}
\rho^{\prime}_{g/e}=\frac{1}{4}(A^{\prime}\pm B^{\prime}),
\end{equation}
where $A^{\prime}$ and $B^{\prime}$ have the same forms as
Eqs.~(\ref{eq:21b}) and (\ref{eq:21c}), just with the replacement
of $\rho_{-}(\tau)$ by $\rho_{+}(\tau)$.

The  probabilities $P^{\prime}_{g}(\tau)$ and
$P^{\prime}_{e}(\tau)$ to measure the qubit states $|g\rangle$ and
$|e\rangle$ corresponding to the prepared state
$|\beta_{+}\rangle$ of Eq.~(\ref{eq:10}) after a dissipation
interval $\tau$, can also be obtained as
\begin{equation}\label{eq:25}
P_{g/e}^{\prime}(\tau)=\frac{1}{2}\left\{1\pm {\rm Re}({\rm
Tr}_{F} [\exp(-i\varphi)\rho_{+}(\tau)])\right\}
\end{equation}
where ${\rm Re}\{{\rm Tr}_{F} [\exp(-i\varphi)\rho_{+}(\tau)]\}$
can be obtained by replacing $N_{-}$ with $N_{+}$, and replacing
the sign $``-"$ before the second and third terms with the sign
$``+"$ in Eq.~(\ref{eq:24}a)

\begin{figure}
\includegraphics[width=42mm]{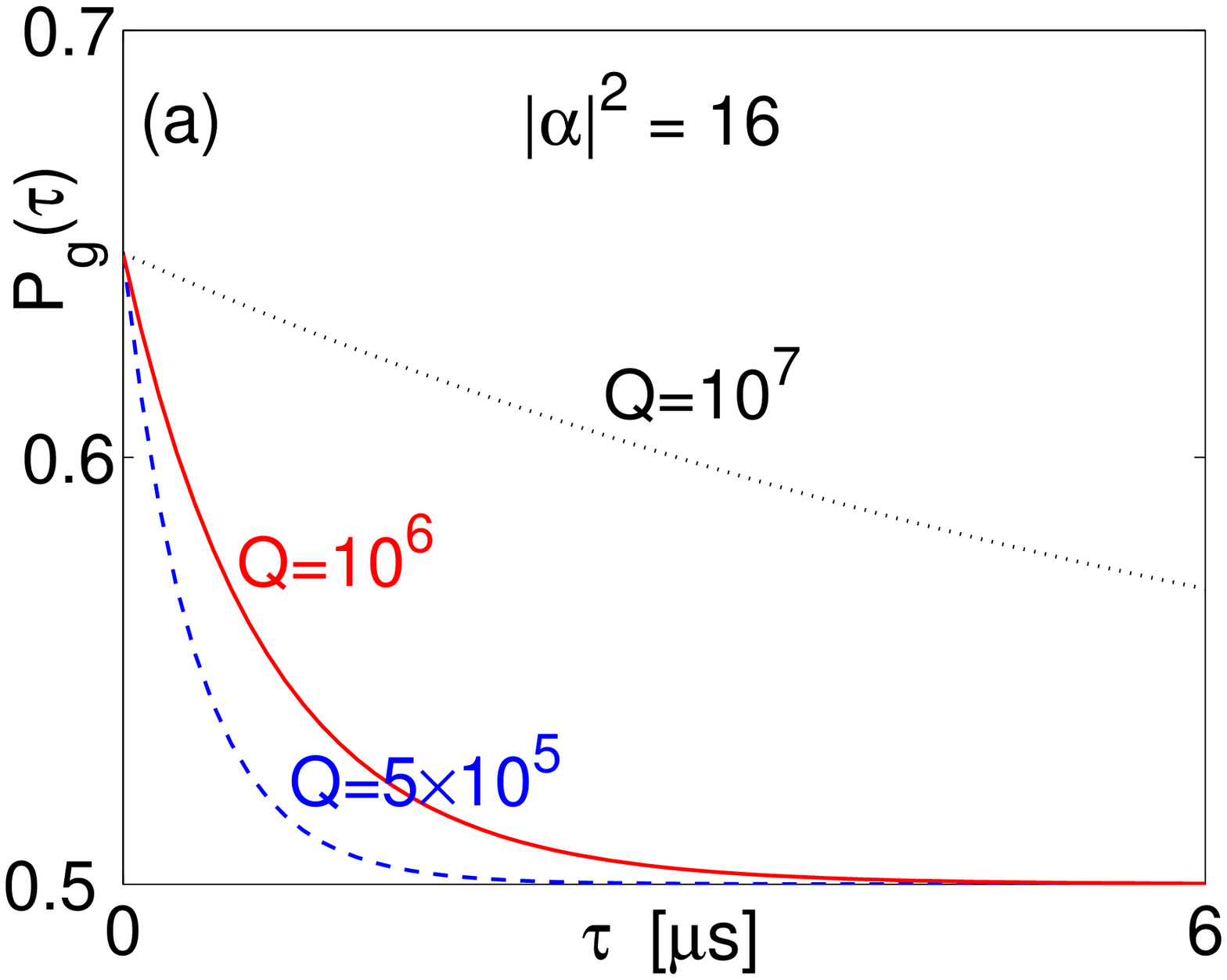}
\includegraphics[width=42mm]{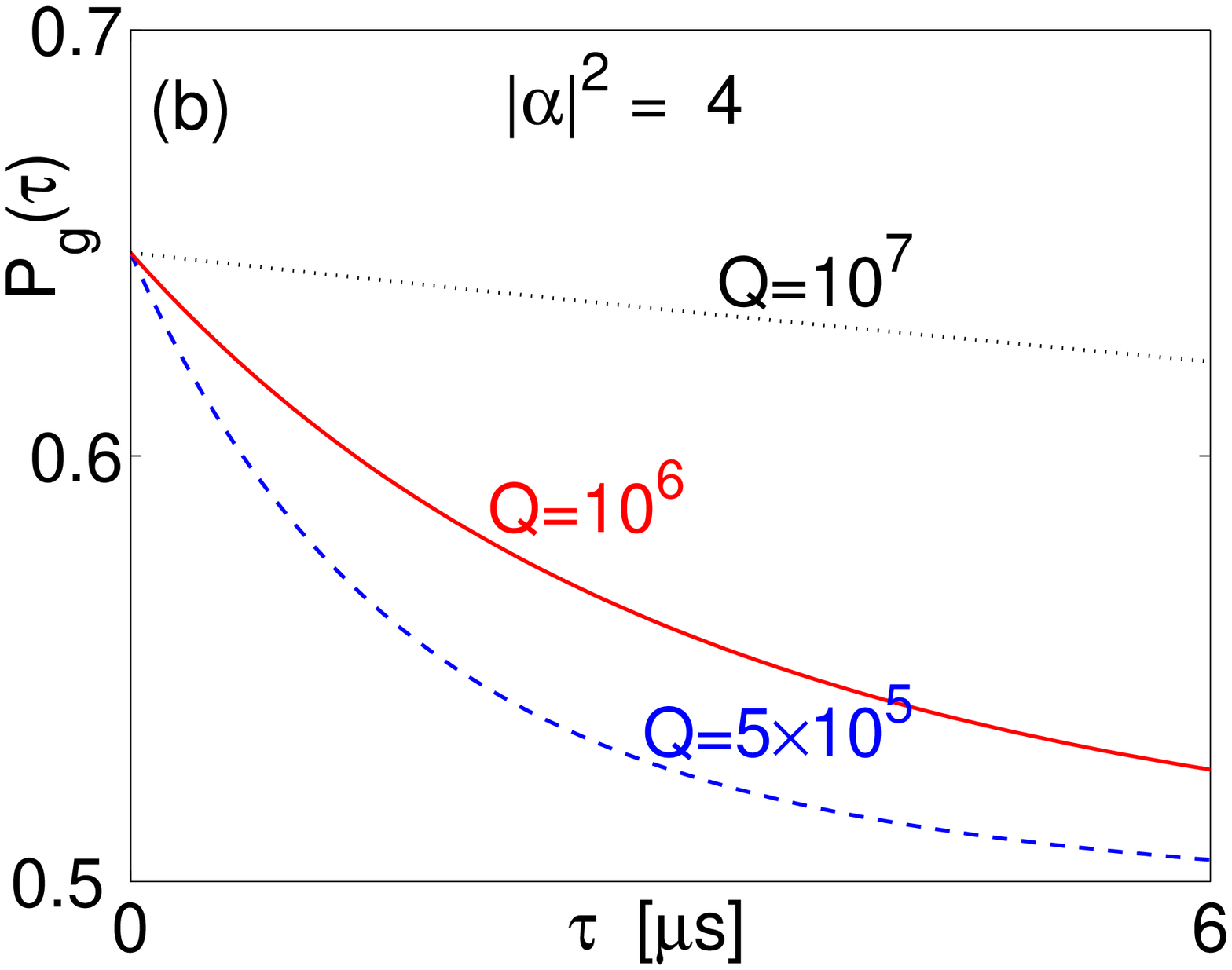}
\caption[]{(Color online) The probability $P_{g}(\tau)$ to measure
the qubit ground state $|g\rangle$ as a function of the evolution
time $\tau$. This $P_{g}(\tau)$ is shown for several values of the
quality factor $Q$ and for different intensities of the input
coherent state $|\alpha\rangle$.}\label{fig2}
\end{figure}

To determine the Q values by probing the charge states, the
measurement should be made for two times, the first measurement is
for the preparation of the superpositions of the cavity field.
After the first measurement, we make a suitable qubit rotation,
and then make the qubit interact with the cavity field for a
duration $\tau_{4}$.  Finally, the second measurement is made and
the $Q$ information is encoded in the measured probabilities. The
different evolution times $\tau$ correspond to the different
measuring probabilities for given $\tau_{4}$ and other parameters
$|\alpha|$, $\Delta$, and so on. For example, the probabilities
$P_{e/g}(\tau)$ for several special cases are discussed as follows
when the prepared state is $|\beta_{-}\rangle$. If we assume that
the qubit rotations and qubit-photon dispersive interaction are
made without energy dissipation of the cavity field, e.g.,
$\tau=0$, then the measuring probabilities $P_{e/g}(\tau=0)$ only
encode the information of the cavity field but do not include the
quality factor $Q$. If the coherence of the states
$|\beta_{\pm}\rangle$ nearly disappears after time $\tau$, then
the state $|\beta_{-}\rangle$ becomes a classical statistical
mixture
\begin{equation}
\rho_{-}(\tau)=\frac{1}{N^2_{-}}\left[|\beta u(\tau)\rangle\langle
\beta u(\tau)|+|\beta^{\prime} u(\tau)\rangle\langle
\beta^{\prime} u(\tau)|\right].
\end{equation}
The probabilities $P_{e/g}(\tau)$ are then reduced to
\begin{eqnarray}
P_{e/g}(\tau)&=&\frac{1}{N^2_{-}}\pm \frac{\exp\left[-2\alpha(\tau)\sin^2\phi^{\prime}\right]}{N^2_{-}} \\
&\times&\cos\left[\Omega_{-}\tau_{4}-\alpha(\tau)\sin(2\phi^{\prime})\right],\nonumber
\end{eqnarray}
which tends to $1/2$ for $|\alpha|^2\gg 1$. If $\tau
>1/\gamma=t_{\rm ph}$ of  the single-photon state lifetime, then the
photons of the states $|\beta_{\pm}\rangle$ are completely
dissipated into the environment. In this case, the cavity quality
factor $Q$ cannot be encoded in the probabilities $P_{e/g}(\tau)$
even with some qubit and qubit-photon states operations.

As an example, let us consider how $P_{g}(\tau)$ varies with the
evolution time $\tau$ with the cavity field dissipation. We assume
that the evolution time $\tau_{4}=(\pi/2) (\Delta/|g|^2)$, that
is, $\phi^{\prime}=\pi/2$. Then, the $\tau$-dependent
probabilities $P_{g}(\tau)$ for the initially prepared state
$|\beta_{-}\rangle$ are given in Fig.~\ref{fig2}(a) with the same
parameters as in Fig.~\ref{fig1}, except with different cavity
quality factors $Q$. In order to see how the probability
$P_{g}(\tau)$ changes with the intensity $|\alpha|^2$ of the input
coherent state $|\alpha\rangle$,  we plot $P_{g}(\tau)$ in
Fig.~\ref{fig2}(b) with the same parameters as Fig.~\ref{fig2}(a)
except changing the intensity to $|\alpha|^2=4$ from
$|\alpha|^2=16$. Figure~\ref{fig2} shows that both the higher
quality factor $Q$ and weaker intensity $|\alpha|^2$ of the input
cavity field correspond to a larger probability $P_{g}$ of the
ground state for the fixed evolution time $\tau$. For fixed $Q$
and $\tau$, the weaker intensity $|\alpha|^2$ corresponds to a
higher measuring probability. We plot $P_{g}(\tau)$ in
Fig.~\ref{fig2} considering the simple case $\phi=\pi$. However,
if we consider another $\phi$, then $|\alpha|^2$ should be chosen
such that it satisfies the condition $2|\alpha|\sin(\phi/2)>1$. In
conclusion, the quality factor $Q$ can be determined from the
probabilities $P_{e/g}(\tau)$ of measuring the qubit states with a
finite cavity field evolution time $\tau$.

\section{Discussions and conclusions}

We discussed how to measure the cavity quality factor $Q$ by using
the interaction between a single-mode microwave cavity field and a
controllable superconducting charge qubit. Two methods are
proposed. One measures the Wigner function of the state
(\ref{eq:11}) by using a standard optical method~\cite{hans}.
Another approach measures the qubit states. Using this last
method, the information of the $Q$ value can be encoded into the
reduced density matrix of the cavity field, and at the same time
the qubit makes a $\pi/2$ rotation. Thus, with a suitable
qubit-photon interaction time, information on the $Q$ value is
then transferred to the qubit-photon states. Finally, after
another $\pi/2$ rotation, the charge qubit states are measured,
and the $Q$ value can be obtained, as shown in Fig.~\ref{fig2},
Eqs.~(\ref{eq:24}) and Eq.~(\ref{eq:188}). However, it should be
noticed that it is easy to measure charge states than to measure
photon states in superconducting circuits.

Our proposal shows that a cavity QED experiment with a SC qubit
can be performed even for a relatively low $Q$ values, e.g.,
$Q\sim 10^{6}$. Initially, a coherent state is injected into the
cavity, which is relatively easy to do experimentally. Although
all rotations of the qubit are chosen as $\pi/2$ to demonstrate
our proposal, other rotations can also be used to achieve our
goal.

To simplify these studies and without loss of generality, we have
assumed two  components $|\beta\rangle$ and $|\beta
\exp(i\phi)\rangle$ for superpositions with $\phi=\pi$ phase
difference in our numerical demonstrations. Of course, other
superpositions can also be used to realize our purpose. The only
condition to satisfy is that the distance between the two states
$|\beta\rangle$ and $|\beta \exp(i\phi)\rangle$ should be larger
than one. In order to obtain a numerical estimate for the
detuning, we specify a value of the gate charge number $n_{g}$.
However, any gate charge that satisfies the large-detuning
condition can be chosen to realize our proposal.

Although we did not give a detailed description of another
resonance-based approach, it should be pointed out that the $Q$
values can also be determined by virtue of the resonant
qubit-photon interaction. For example, if the
superpositions~\cite{liu} of the vacuum and the single photon
state are experimentally prepared, then we can follow the same
steps as in Sec.~III and IV to obtain the $Q$ value. This
method~\cite{rinner} has been applied to micromasers, where the
qubits are two-level atoms. However, the coherent states and
non-resonant qubit-photon interaction should be easier to do
experimentally than the approach using single-photon states and
resonant qubit-photon interaction.

Our proposal can also be generalized to the models used in
Refs.~\cite{wallraff,blais}, which are experimentally accessible.
We hope that our proposal can open new doors to experimentally
test the $Q$ value and motivate further experiments on  cavity
quantum electrodynamics with SC qubits.

\section{acknowledgments}

This work was supported in part by the National Security Agency
(NSA) and Advanced Research and Development Activity (ARDA) under
Air Force Office of Research (AFOSR) contract number
F49620-02-1-0334, and by the National Science Foundation grant No.
EIA-0130383.

\appendix
\section{Effective hamiltonian with larger detuning}

The Hamiltonian $H=H_{0}+H_{1}$ of the two-level atom interacting
with a single-mode cavity field can be written as
\begin{subequations}
\begin{eqnarray}
H_{0}&=&\frac{1}{2}\hbar\Omega\sigma_{z}+\hbar\omega
a^{\dagger }a,\\
 H_{1}&=&\hbar (g  a^{\dagger }\sigma_{-}+g^{*} a\sigma_{+})
\end{eqnarray}
\end{subequations}
with a complex number $g$. Let us assume $\Delta=\Omega-\omega >
0$ and $g/\Delta \ll 1$. The eigenstates and corresponding
eigenvalues of the free Hamiltonian $H_{0}$ are
\begin{subequations}
\begin{eqnarray}
|e\rangle\otimes|n\rangle&\Longrightarrow&
n\hbar\omega+\frac{1}{2}\hbar\Omega,\\
|g\rangle\otimes|m\rangle&\Longrightarrow&
m\hbar\omega-\frac{1}{2}\hbar\Omega
\end{eqnarray}
\end{subequations}
In the interaction picture,  any state can be written as
\begin{equation}
|\psi(t)\rangle=U(t,t_{0})|\psi(t_{0})\rangle
\end{equation}
with
\begin{eqnarray}\label{a4}
U(t,t_{0})&=&1+\frac{1}{i\hbar}\int_{t_{0}}^{t}H_{\rm
int}(t_{1}){\rm
d}t_{1}\\
&+&\left(\frac{1}{i\hbar}\right)^2\int_{t_{0}}^{t}\int_{t_{0}}^{t_{1}}H_{\rm
int}(t_{1})H_{\rm int}(t_{2}){\rm d}t_{1}{\rm d}t_{2}+\cdots
\nonumber
\end{eqnarray}
here $H_{\rm int}=U^{\dagger}_{0}(t)H_{1}U_{0}(t)$ with
$U_{0}(t)=\exp\{-iH_{0}t/\hbar\}$. In the basis
$\{|E_{l}\rangle\}=\{|e\rangle\otimes|n\rangle,\,\,|g\rangle\otimes|m\rangle\}$,
Eq.~(\ref{a4}) can be expressed as
\begin{eqnarray}
U(t,t_{0})&=&1+\nonumber\\
&+&\frac{1}{i\hbar}\int_{t_{0}}^{t}\sum_{l,m}|E_{l}\rangle\langle
E_{l}|H_{\rm int}(t_{1})|E_{m}\rangle\langle E_{m}|{\rm
d}t_{1}+\cdots. \nonumber
\end{eqnarray}
After neglecting the fast-oscillating factor and keeping the first
order term in $g/\Delta$, $U(t,t_{0})$
\begin{eqnarray}
&&U(t,t_{0})=U(t,0)=U(t)\\
&&\approx 1-i\frac{|g|^2}{\Delta}\int_{0}^{t}{\rm
d}t_{1}[(n+1)|e,n\rangle\langle e,n|-n|g,n\rangle\langle
g,n|]\nonumber,
\end{eqnarray}
where we assume $t_{0}=0$. Finally, we obtain the effective
Hamiltonian in the interaction picture as
\begin{equation}\label{a6}
H_{\rm eff}=\hbar\frac{|g|^2}{\Delta}(|e\rangle \langle
e|aa^{\dagger}-|g\rangle \langle g|a^{\dagger}a).
\end{equation}
Returning Eq.~(\ref{a6}) to the Schr\"odinger picture,
Eq.~(\ref{eq:6}) is obtained. This method can be generalized to
obtain the effective Hamiltonian of the model with many two-level
system interacting with a common single-mode field.
Equation~(\ref{eq:6}) can also be obtained by using the
Fr\"ohlich-Nakajima transformation~\cite{Fro,Nakajima,wu,sun}.

\section{Wigner functions of superposition and mixed states}

For completeness, we explicitly write the Wigner functions
$W_{\pm}(x,p)$ of the superposition states in Eq.~(\ref{eq:10}) as
follows:
\begin{eqnarray}
&&W_{\pm}(x,p)\nonumber\\
&& =\frac{1}{\pi N^{2}_{\pm}}\left\{\exp
\left[-\left(x-\sqrt{2}\,{\rm Re}\beta
\right)^2-\left(p-\sqrt{2}\,{\rm Im}\beta
\right)^2\right]\right.\nonumber\\
&&+\exp \left[-\left(x-\sqrt{2}\,{\rm Re}\beta^{\prime}
\right)^2-\left(p-\sqrt{2}\,{\rm Im}\beta^{\prime}
\right)^2\right]\nonumber\\
&&\left.\pm 2 {\rm Re}\left[P \exp \left(-\left(x-\wp_{1}
\right)^2-\left(p+i\wp_{2} \right)^2\right)\right]\right\}\, ,
\end{eqnarray}
with
\begin{eqnarray}
P&=&\exp(-i\theta)\exp[-|\alpha|^2(1-e^{i\phi})],\\
\wp_{1}&=&\frac{1}{\sqrt{2}}(\beta+\beta^{\prime*}),\\
\wp_{2}&=&\frac{1}{\sqrt{2}}(\beta-\beta^{\prime*}).
\end{eqnarray}
The Wigner functions $W_{\pm D}(x,p)$ of the mixed states in
Eq.~(\ref{eq:11}) with dissipation can be written as
\begin{eqnarray}
&&W_{\pm D}(x,p)\nonumber\\
&& =\frac{1}{\pi N^{2}_{\pm}}\left\{\exp
\left[-\left(x-u\sqrt{2}\,{\rm Re}\beta
\right)^2-\left(p-u\sqrt{2}\,{\rm Im}\beta
\right)^2\right]\right.\nonumber\\
&&+\exp \left[-\left(x-u\sqrt{2}\,{\rm Re}\beta^{\prime}
\right)^2-\left(p-u\sqrt{2}\,{\rm
Im}\beta^{\prime} \right)^2\right]\nonumber\\
&&\left.\pm 2 {\rm Re}\left[P \exp \left(-\left(x-u\,\wp_{1}
\right)^2-\left(p+i u \,\wp_{2} \right)^2\right)\right]\right\}.
\end{eqnarray}

\end{document}